\documentclass[aps,twocolumn,pra,tightenlines,showpacs]{revtex4-1}
\usepackage[dvips]{graphicx}
\usepackage[english]{babel}
\usepackage{amsmath}
\usepackage{amssymb}
\usepackage{times}
\def\be{\begin{equation}}
\def\ee{\end{equation}}
\def\bea{\begin{eqnarray}}
\def\eea{\end{eqnarray}}

\newcommand{\ek}{\xi_{\mathbf{k}}}
\newcommand{\Ek}{{E^{}_{\mathbf{k}}}}

\newcommand{\mb}[1]{{\mathbf{#1}}}

\renewcommand{\text}[1]{{#1}}

\begin{document}

\title{Probing the homogeneous spectral function of a strongly
  interacting superfluid atomic Fermi gas in a trap using phase
  separation and momentum resolved rf spectroscopy}

\author{Qijin Chen}

\affiliation{Department of Physics and Zhejiang Institute of Modern
  Physics, Zhejiang University, Hangzhou, Zhejiang 310027, CHINA}

\date{\today}

\begin{abstract}
  It is of central importance to probe the \emph{local} spectral
  function $A(\mathbf{k},\omega)$ of a strongly interacting Fermi gas
  in a trap.  Momentum resolved rf spectroscopy has been demonstrated
  to be able to probe the trap averaged
  $A(\mathbf{k},\omega)$. However, the usefulness of this technique
  was limited by the trap inhomogeneity. Independent of a specific
  theory, here we propose that by studying the momentum resolved rf
  spectra of the minority fermions of a phase separated, population
  imbalanced Fermi gas at low temperature, one can effectively extract
  $A(\mathbf{k},\omega)$ of a homogeneous superfluid Fermi gas (at the
  trap center).  In support, we present calculated spectral functions
  and spectral intensity maps for various cases from BCS through BEC
  regimes using different theories.
\end{abstract}


\maketitle

\section{Introduction}

Ultracold atomic Fermi gases has emerged as a rapidly developing
field, bridging condensed matter and atomic physics. 
Owing to their easy tunabilities such as the effective two-body
interaction strength, population imbalance, and dimensionality, atomic
Fermi gases can be viewed as a quantum simulator of many important
condensed matter systems, such as the Hubbard model and high Tc
superconductors \cite{LeggettNature,ourreview}.

Of central importance in a many-body system is the single particle
spectral function $A(\mathbf{k},\omega)$, especially in the strongly
interacting regime (where different theories often do not agree); 
it can be used to calculate essentially all other physical quantities,
and to test various theories. For example, one important question,
which has been long standing for high Tc superconductors, is whether a
pseudogap exists and how it evolves in the superfluid state.  Because
measurements often involve integration of the entire momentum space
and the entire trap, the intrinsic spatial inhomogeneity introduced by
the trapping potential constitutes a severe problem.
It has been a long-sought goal to extract the \emph{local or
  homogeneous} properties, especially $A(\mathbf{k},\omega)$, from
measurements in the trap.

In this paper, we propose that one can use momentum resolved rf
spectroscopy in conjunction with a high population imbalance to
measure the homogeneous spectral function $A(\mathbf{k},\omega)$, for
the density and chemical potential at the trap center.

Since the first experimental realization of BCS-BEC crossover, the
most direct experimental probe of the single particle properties is
arguably the rf spectroscopy. The first generation of rf measurements
involves integration over the entire trap and the whole momentum space
\cite{Grimm4}, which led to a two-peak structure of the rf spectra at
an intermediate temperature $T$ below $T_c$.  Due to the integrations,
controversies arose regarding the origin of this two-peak structure.
Tomographic rf spectroscopy \cite{Tomography} involves integration
over the entire momentum space, and therefore cannot be used to probe
the spectral function.
The recent Ho-Zhou scheme \cite{HoZhou_nphys} allows one to calculate
the density and fermionic chemical potential $\mu$ as a function of
the radius $r$. However, it relies on the assumption of a \emph{strict}
harmonic trapping potential \cite{HoPrivate}, often not satisfied.
More importantly, it does not provide any microscopic information such
as the spectral function.

The controversies regarding the origin of the aforementioned double
peak structure have largely been cleared by the recent experiments
\cite{Jin6} on $^{40}$K using momentum resolved rf spectroscopy. With
momentum resolution, the rf spectroscopy essentially measures the
centrally important $A(\mathbf{k},\omega)$, which is of central
importance in a many-body system.
However, this is true only for a \emph{homogeneous} Fermi gas.  The
inherent spatial inhomogeneity severely limits the quantitative
resolution of the extracted spectral function in the experiment of
Ref.~\cite{Jin6}, which involves integration over the entire trap.

Here we address this inhomogeneity issue and show that it can be
largely avoided utilizing phase separation at high population
imbalances.  In what follows we will use two different theories to do
computations. Despite some differences in \emph{quantitative details},
we emphasize that \emph{the validity of our proposal is model
  independent} and can be applied in other theories as well. Most
importantly, \emph{it is a proposal for experiment}.  

As mentioned elsewhere \cite{Chen_MRRF}, the use of $^{40}$K rather
than $^6$Li is crucial in extracting $A(\mathbf{k},\omega)$, since
there are no complications from final state interactions
\cite{FinalStateEffects} near the usual Feshbach resonance around
202~G.  It is for this reason that the rf spectral intensity for
$^{40}$K is simply proportional to the spectral function
$A(\mathbf{k},\omega)$. For $^6$Li, extra efforts are needed in order
to extract $A(\mathbf{k},\omega)$ from the rf spectral measurements.

The spectral function $A(\mathbf{k},\omega)$ can be used to uniquely
construct the Green's function and the single particle self energy,
for which different theories often give different results. 
 Therefore, the more accurate the measurement of
 $A(\mathbf{k},\omega)$, the easier it is to test different theories.
For example, different BCS-BEC crossover approaches
\cite{Levin_AnnPhys_RPP,SuSQ} show different dispersive behavior of
$A(\mathbf{k},\omega)$. The theoretical scheme used in
Ref.~\onlinecite{Chen_MRRF} exhibits a clear downward dispersion in
the spectral intensity map near $T_c$, manifesting an existing
pseudogap before the onset of superfluidity at lower $T=T_c$,
different from those \cite{Strinaticuprates,Stoof3} that follow the
approach of Nozi\'eres and Schmitt-Rink (NSR) \cite{NSR}.
A \emph{quantitatively accurate} measurement of $A(\mathbf{k},\omega)$
should serve to unambiguously test these different schools of
theories.

\section{Theoretical Formalism}

In a typical rf spectroscopy measurement, pairing takes place between
two low-lying hyperfine states, which we refer to as levels 1 and 2.
An rf field 
of frequency $\nu$
is used to excite the atoms in hyperfine level 2 to another hyperfine
state, which is unoccupied initially and referred to as level 3.  It
has been shown previously \cite{Chen_MRRF} that for $^{40}$K
%
we have for the momentum-resolved RF current
\begin{eqnarray}
I(\mathbf{k}, \delta\nu)
  &=&\left. \frac{1}{2\pi} A(\mathbf{k}, \omega)
  f(\omega)\right|_{\omega=\ek -\delta\nu} ,
\label{eq:RFcurrent}
\end{eqnarray}
where $A(\mb{k},\omega) =-2\,\mbox{Im}\, G(\mb{k},\omega+i0)$, $f(x)$
the Fermi distribution function, 
with the transition matrix element set to unity. Here $G$ is the
Green's function of level 2 atoms, which we take to be the spin down
or minority species.  Similar to angle-resolved photoemission
spectroscopy in a usual condensed matter system, the rf current
measures $A(\mb{k},\omega)$ directly.
Note that here $\delta\nu$ is the rf detuning.  The frequency
$\omega=\ek -\delta\nu$ corresponds to the energy of level 2 atoms
measured with respect to their Fermi level, where $\ek = k^2/2m -\mu$,
$\mu$ is the chemical potential of atoms in level 2.  We take $\hbar
=1$. It is obvious that when level 2 atoms are free, we have
$\delta\nu=0$ and $\omega=\ek$; the former gives the sharp peak at
zero detuning in previous rf spectra in Ref.~\onlinecite{Grimm4}
whereas the latter gives the free atom dispersion in the spectral
intensity map in the $\omega-k$ plane in Ref.~\onlinecite{Jin6}.
As has been used in the experiment \cite{Jin6}, the angle-integrated
``occupied spectral intensity'' is given by
$I^{photo}(k, \omega)  \equiv (k^2/2\pi^2) A(k,\omega)f(\omega)$.

The central issue here is to calculate the spectral function or
equivalently the Green's function $G(K)$ for the minority (level 2)
atoms. Here we present calculations using two different approaches. In
the first, \emph{pairing fluctuation}, approach, in the presence of
population imbalance, detailed calculations of the superfluid phase
diagram, the Green's function and density profiles in each phase can
be found in Refs.~\onlinecite{ChienPRL,LOFF1_HePRB2007}. In the case
of phase separation in a trap, as shown in Fig.~2 of
Ref.~\onlinecite{ChienPRL}, there is a BCS superfluid core (without
population imbalance), surrounded by the majority atoms, as confirmed
experimentally \cite{ZSSK06_Rice1,Shin08}. To a first order
approximation, the density profile of the minority atoms can be
obtained by truncating the density profile of an unimbalanced Fermi
gas at the (sharp) phase separation boundary.
With and without population imbalance, the normal state self energy
$\Sigma_{pg}$ follows a rather simple BCS-like form
\cite{Chen4,Chen_MRRF}. In the superfluid state, the self
energy $\Sigma(K)$ contains two terms, associated with the condensed
($\Sigma_{sc}$) and noncondensed fermion pairs ($\Sigma_{pg}$),
respectively: $\Sigma({\bf k}, \omega) = \Sigma_{pg}({\bf k}, \omega )
+ \Sigma_{sc}({\bf k}, \omega )$, where
\begin{equation} \Sigma_{pg}(\mb{k},\omega) = 
  \frac{\Delta_{pg}^2}{\omega+\ek+i\gamma} -i\Sigma_0, 
\quad 
\Sigma_{sc}(\mb{k},\omega) =
\frac{\Delta_{sc}^2}{\omega +\ek} \,.
  \label{eq:Sigma_1}
\end{equation}
Here $\Delta_{pg}$, $\Delta_{sc}$ and
$\Delta=\sqrt{\Delta_{sc}^2+\Delta_{pg}^2}$ are the pseudgogap,
superfluid gap and total excitation gap, respectively. The broadening
$\gamma \ne 0$ and ``incoherent'' background contribution $\Sigma_0$
can be determined by fitting the experimentally measured rf
spectra. The precise values of $\gamma$ and $\Sigma_0$, and their
$T$-dependencies are not particularly important for the present
purposes. Although the pseudogap contribution becomes less pronounced
when the temperature is low enough to exhibit phase separation, it
still makes a significant difference in comparison with a strict BCS
theory.

In the second, \emph{mean-field}, approach, as a test case, we use the
same density profiles as obtained from the first approach, but take a
simple broadened BCS self energy,
\begin{equation} 
 \Sigma_{BCS}(\mb{k},\omega) = 
  \frac{\Delta^2}{\omega+\ek+i\gamma} -i\Sigma_0.
  \label{eq:Sigma_2}
\end{equation}
Both approaches give the same form of quasiparticle dispersion $\Ek =
\sqrt{\xi_\mathbf{k}^2 + \Delta^2}$ albeit with different meanings of
$\Delta$.  Obviously, they become equivalent above $T_c$, when we take
$\gamma = \Sigma_0$.  In contrast, they differ dramatically below
$T_c$ \cite{Chen4} because one contains a pseudogap while the other
does not. In both approaches, $\gamma$ and $\Sigma_0$ scale with local
$E_F(r)$. As in Ref.~\onlinecite{Chen_MRRF}, here we will take
$\Sigma_0$ as $T$ independent, $\gamma$ linear in $T/T_c$, where $T_c$
is calculated using the first approach. The rf spectra are finally
convoluted with a Gaussian broadening function with a standard
deviation $\sigma$, reflecting the instrumental resolution in
experiment caused by e.g. finite energy and momentum resolution of
both the rf pulse and the time of flight imaging technique.

\begin{figure}[tb]
\centerline{\includegraphics[width=3.in,clip]{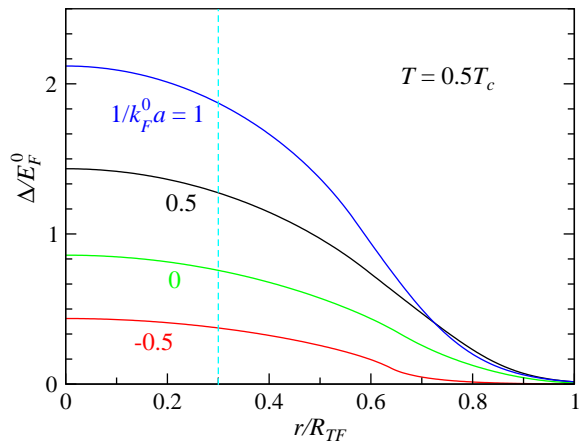}}
\caption{(Color online) Radial profile of the excitation gap
  $\Delta/E_F^0$ in the superfluid phase of a trapped Fermi gas
  without population imbalance for $1/k_F^0a = -0.5, 0, 0.5$ and 1,
  corresponding to BCS through BEC regimes, calculated at $T= 0.5T_c$,
  using self energy Eq.~(\ref{eq:Sigma_1}). The gap is nearly flat for
  $r<0.3R_{TF}$.}
\label{fig:2}
\end{figure}

\section{Results and Discussions}

We first study the radial profile of the excitation gap 
in the superfluid phase. While throughout this paper we study each
case in a harmonic trap with a local density approximation using both
approaches, here we show in Fig.~\ref{fig:2} the result from the
\emph{pairing fluctuation approach} Eq.~(\ref{eq:Sigma_1}).  It
suffices to calculate for cases without population imbalance, since
upon phase separation the superfluid core becomes unimbalanced so that
the gap profile of minority atoms can be obtained by cutting off the
curves in Fig.~\ref{fig:2} at different radii for different population
imbalances $p\equiv (N_\uparrow - N_\downarrow)/(N_\uparrow +
N_\downarrow)$. For simplicity, here the data were all calculated at
$T=0.5T_c$, below which phase separation takes place at unitarity
\cite{ChienPRL}. The curves, as labeled, correspond to $1/k_F^0a =
-0.5 $ (BCS case), 0 (unitary), 0.5 (pseudogap) and 1.0 (BEC) cases,
respectively, where $a$ the inter-fermion $s$-wave scattering length,
$E_F^0\equiv k_BT_F^0 \equiv \hbar^2(k_F^0)^2/2m$ and $k_F^0$ are the
global Fermi energy and Fermi momentum in the noninteracting limit for
the majority atoms.  Figure \ref{fig:2} reveals that for
$r<0.3R_{TF}$, the gap is nearly flat from BCS through BEC regimes,
where $R_{TF}$
is the Thomas-Fermi radius. Since the pseudogap increases with pairing
strength and it is already comparable with the zero $T$ gap at
unitarity \cite{ourreview,Jin6}, the \emph{total} gap profile
$\Delta(r)$ remains nearly flat for small $r$ even at higher $T\approx
T_c$ at unitarity and in the BEC regime \cite{BECProfileFootnote}.

\begin{figure}
\centerline{
\includegraphics[width=3.4in,clip]{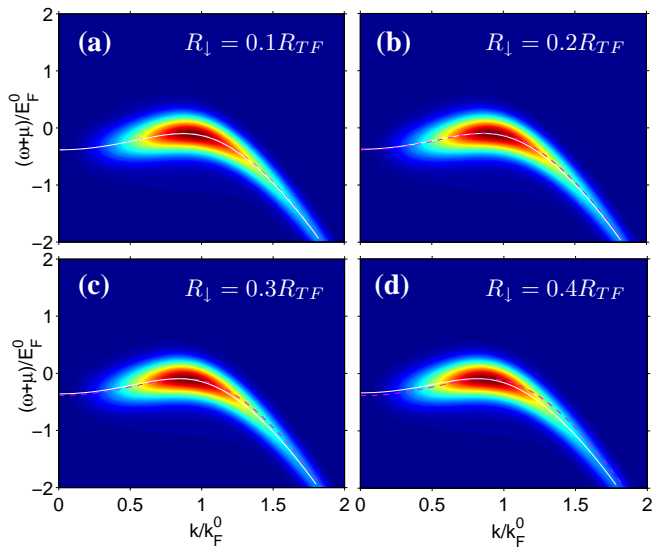}}
\caption{(Color online) Occupied spectral intensity maps
  $I^{photo}(k,\omega)$ for the minority atoms of a population
  imbalanced, phase separated unitary Fermi gas in a harmonic trap for
  (a) $R_\downarrow = 0.1R_{TF}$, (b) $0.2R_{TF}$, (c) $0.3R_{TF}$,
  and (d) $0.4R_{TF}$, respectively, calculated at $T=0.13T_F^0$,
  using self energy Eq.~(\ref{eq:Sigma_1}).   The
  (magenta) dashed curves represent the local (or homogeneous)
  dispersion at the trap center, and the (white) solid curves are the
  quasiparticle dispersion given by the loci of the peak location of
  the EDC. The broadening parameters 
  $\Sigma_0 = \gamma(T_c) = 0.1 E_F^0$ at the trap center. Here the
  spectral intensity increases from 0 (dark blue) continuously through
  the maximum (dark red). The quantitative scale for, e.g., panel (c)
  may be seen from Fig.~\ref{fig:5}. For the instrumental Gaussian
  broadening, we take $\sigma = 0.2E_F^0$ (as extracted
  \protect\cite{Chen_MRRF} from the experiment in
  Ref.~\protect\onlinecite{Jin6}).  Up to $R_\downarrow= 0.4R_{TF}$,
  the two sets of curves are essentially indistinguishable.}
\label{fig:3}
\end{figure}
		 
With this knowledge, now we study the occupied spectral intensity maps
of the minority (spin down) atoms in the $(\omega+\mu(r))$ -- $k$
plane for different degrees of population imbalance $p$, as
characterized by the radius $R_\downarrow$ of the minority atomic
gas. For illustration purpose, we present the result from the
\emph{pairing fluctuation approach} Eq.~(\ref{eq:Sigma_1}) for the
unitary case
with phase separation in Fig.~\ref{fig:3}, for different
$R_\downarrow$.
Note here that, as in Ref.~\onlinecite{Chen_MRRF}, we use
$\omega+\mu(r)$ instead of $\omega$ in the vertical axis, since the
former combination is independent of $r$ representing the single
particle energy measured from the bottom of the band. While
Fig.~\ref{fig:3}(a) corresponds to an extremely high population
imbalance, which may not be readily realizable in experiment,
Fig.~\ref{fig:3}(c)-(d) are certainly accessible experimentally
\cite{ZSSK06_Rice1}. 
For example, for $^6$Li, Ref.~\cite{Shin08} reports a phase separation
boundary at about $0.3 R_{TF}$ at $p = 0.54$.
The (magenta) dashed curves represent the local (or homogeneous)
dispersion at the trap center, and the (white) solid curves are the
quasiparticle dispersion given by the loci of the peak location of the
energy distribution curves (EDC). It is evident that up to
$R_\downarrow= 0.4R_{TF}$, the two sets of curves are essentially
indistinguishable.

\begin{figure}
\centerline{
\includegraphics[width=3.4in,clip]{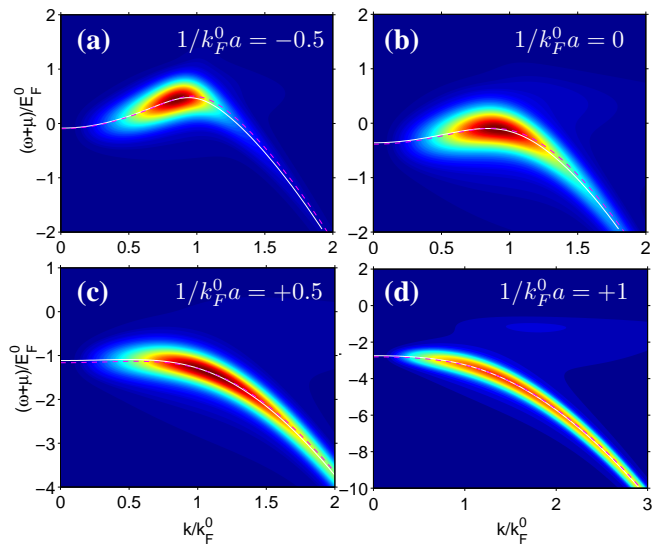}}
\caption{(Color online) Occupied spectral intensity maps
  $I^{photo}(k,\omega)$ for the minority atoms of a population
  imbalanced, phase separated Fermi gas with $R_\downarrow =
  0.3R_{TF}$ for (a) $1/k_F^0a = -0.5$ (BCS case), (b) 0 (unitary),
  (c) 0.5 (pseudogap), and (d) 1.0 (BEC case), calculated at
  $T=0.09T_F^0$, $0.13T_F^0$, $0.15T_F^0$ and $0.15T_F^0$,
  respectively, \emph{assuming self energy} Eq.~(\ref{eq:Sigma_2}).
  The legends are the same as in Fig.~\ref{fig:3}.  The parameters
  $\Sigma_0 = \gamma(T_c)$ at the trap center are (a) $0.1 E_F^0$,
  (b)-(c) $0.2 E_F^0$, and (d) $0.35E_F^0$, respectively. $T_c$ is
  taken from the pairing fluctuation approach. The instrumental
  broadening $\sigma = 0.2E_F^0$. For all cases, the two sets of
  curves are essentially indistinguishable.}
\label{fig:4}
\end{figure}

Shown in Fig.~\ref{fig:4} are the occupied spectral intensity maps of
the minority atoms of population imbalanced, phase separated atomic
Fermi gases 
 in a harmonic trap
with 
 minority radius
$R_\downarrow = 0.3R_{TF}$ for different pairing strength from BCS
through BEC.  Here we show the results calculated using the \emph{mean
  field} self energy Eq.~(\ref{eq:Sigma_1}).  Since the atomic cloud
shrinks with increasing $1/k_F^0a$, these same $R_\downarrow$
corresponds to an imbalance of about $p
=0.9 \sim 0.7$. Here the parameters $\gamma$ and $\Sigma_0$ increase
from BCS to BEC, reflecting an increasing excitation gap. Clearly, for
all cases shown, the dispersions extracted from the trap averaged
minority atoms are very close to that calculated at the trap center.

\begin{figure}[t]
\centerline{
\includegraphics[width=3.in,clip]{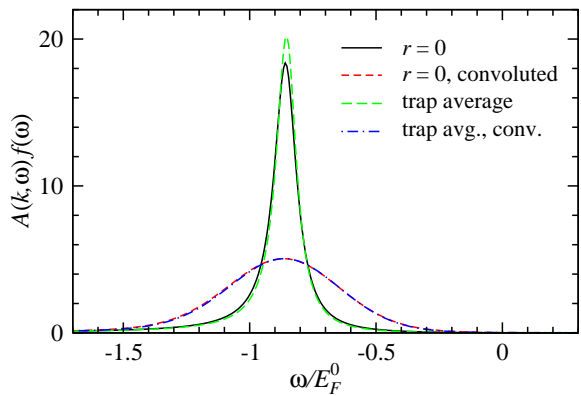}}
\caption{(Color online) Comparison of occupied spectral function
  $A(k,\omega)f(\omega)$ between atoms at the trap center (black solid
  and red dashed curves) and trap averaged minority atoms (green long
  dashed and blue dot dashed curves), for a population imbalanced,
  phase separated unitary Fermi gas with $R_\downarrow = 0.3R_{TF}$ at
  $T=0.5T_c=0.13T_F^0$. The spectra are calculated for
  $k=0.88k_F^0\approx \sqrt{2m\mu(r=0)}$ near the Fermi surface, using
  self energy Eq.~(\ref{eq:Sigma_1}).  The narrow peaks are the
  intrinsic spectral line, whereas the broad ones are those convoluted
  with a instrumental broadening function.  The broadening parameters
  $\Sigma_0$, $\gamma$ and $\sigma$ are the same as in
  Fig.~\ref{fig:2}. The trap averaged minority spectral function is
  nearly the same as that at the trap center. }
\label{fig:5}
\end{figure}

Note that the upward dispersing branch of thermally excited
quasiparticles is completely suppressed by $f(\omega)$ and becomes
invisible here. However, its leftover at large $k$ form a very weak
broad peak at negative $\omega$, as indicated by the light blue area
above the downward dispersing branch. At higher $T$, this feature
\cite{ChenMM09}
 naturally explains the artificial abrupt jump in the
quasiparticle dispersions extracted at higher $T$ using a \emph{single
  peak fit} in Fig.~1 of Ref.~\onlinecite{JinStrinati_nphys}.

Finally, shown in Fig.~\ref{fig:5} is a direct comparison of the
(occupied) spectral function between the atoms at the trap center
(black solid and red dashed curves) and the trap averaged minority
atoms (green long dashed and blue dot-dashed curves) of a highly
population imbalanced, phase separated Fermi gas, from the pairing
fluctuation approach.  The curves are for $k=0.88k_F^0\approx
\sqrt{2m\mu(r=0)}$, which is near the Fermi level [where $A(k,\omega)$
is most sensitive to the variation of $\Delta$] at the trap
center. The spectral peaks at $+\Ek \approx \Delta$ are suppressed by
$f(\omega)$.  The narrow peaks are the spectral lines without
instrumental broadening, and the broad ones are the same result but
with instrumental broadening.  There are small differences between the
unconvoluted spectral peaks (black solid) and (green long dashed
curves) in the peak location and peak height (or equivalently peak
width). This is primarily due to the gradually decreasing broadening
parameters caused by decreasing local $E_F(r)$ as $r$ increases,
leading to slightly higher peak and smaller average gap (given by the
peak location). We emphasize that these differences are barely
discernible and are completely washed out by instrumental broadening,
as revealed by the complete overlap between the two broad peaks. The
complete spectral function $A(k,\omega)$ can be obtained by dividing
the unconvoluted lines by $f(\omega)$.

Indeed, a comparison between Fig.~\ref{fig:3}(c) and
Fig.~\ref{fig:4}(b) reveals that the spectral lines from self energy
Eq.~(\ref{eq:Sigma_2}) have a broader background than that from
Eq.~(\ref{eq:Sigma_1}). Precise spectral function measurement as shown
in Fig.~\ref{fig:5} is expected to tell them apart.

\section{Conclusions}

In conclusion, in the presence of phase separation for a highly
population imbalanced, strongly interacting Fermi gas, the trap
averaged quasiparticle dispersion of the minority atoms is very close
to that at the trap center. Therefore, one can use the former
dispersion to extract effectively not only the excitation gap and the
fermionic chemical potential but also the \emph{centrally important}
spectral function at the trap center. This scheme does not depend on
our particular BCS-BEC crossover theory, nor does it depend on the
strict harmonicity of the trapping potential.  For higher $T$ or
deeper BEC regime where phase separation is prohibited, one may use an
rf pulse with a narrow cross section (as has been reported
experimentally \cite{JinAPS11}) \emph{without} population imbalance to
achieve similar results. In this way, the narrow rf beam mimics the
effect of \emph{artificial} phase separation by picking up signals
only from the central part of the trap \cite{narrowbeamfootnote}.

\acknowledgments

This work is supported by NSF of China (grant No. 10974173), the 973
Program of China (grant No. 2011CB921300), Fundamental Research Funds
for Central Universities of China (Program No. 2010QNA3026), PCSIRT
(Contract No. IRT0754), and by Zhejiang University (grant
No. 2009QNA3015).

\bibliographystyle{apsrev4-1}

\begin{thebibliography}{24}
\bibitem [{BEC()}]{BECProfileFootnote}%
  \BibitemOpen
  \href@noop {} {}\bibinfo {note} {In the BEC regime, the gas cloud shrinks so
  that phase separation boundary also moves torward the trap
  center.}\BibitemShut {Stop}%
\bibitem [{nar()}]{narrowbeamfootnote}%
  \BibitemOpen
  \href@noop {} {}\bibinfo {note} {It should be noted that the boundary of a
  focused rf pulse may not be as sharp as the interface of phase separation,
  making our phase-separation method a better choice at low $T$.}\BibitemShut
  {Stop}%
\end{thebibliography}
%

\end{document}